\documentclass[preprint,tightenlines,showpacs,preprintnumbers,amsmath,amssymb,prb]{revtex4}

\usepackage{graphicx}
\usepackage{bm}
\makeatletter
\def\frontmatter@thefootnote{%
   \altaffilletter@sw{\@alph}{\@fnsymbol}\c@footnote}
\def\frontmatter@makefnmark{%
   \@textsuperscript{\normalfont\@thefnmark)}}
\makeatother

\begin{document}
\draft
\title{Structural transformations of double-walled carbon nanotube bundle under hydrostatic pressure}

\author{Xiaoping Yang}
\altaffiliation[Also at: ]{Department of Physics, Huainan Normal
University, Huainan, Anhui 232001, P. R. China}
\email{bunnyxp@hotmail.com}

\author{Gang Wu}
\altaffiliation[Present address: ]{Department of Physics, National
University of Singapore, Singapore 117542}

\author{Jinming Dong}
\email[Corresponding author. Email address: ]{jdong@nju.edu.cn}

\affiliation{National Laboratory of Solid State Microstructures
and Department of Physics, Nanjing University, Nanjing 210093, P.
R. China}

\begin{abstract}
Three kinds of the response mechanisms to the external pressure
have been found for double-walled carbon nanotube (DWNT) bundle,
depending strongly on their average radius and symmetry. The
small-diameter DWNT bundle undergoes a small discontinuous volume
change, and then deform continuously. The intermediate-diameter
DWNT bundle collapses completely after a structure phase
transition (SPT). Significantly, two SPTs exist for the
larger-diameter DWNT bundle if the outer tube has no $C_{6}$ or
$C_{3}$ symmetry. It would be interesting to search for signatures
of these different structural transformations by experimentally
investigating mechanical, optical and thermal response functions
of DWNT bundle.

\end{abstract}


\date{\today}

\maketitle

It is well known that the physical properties of the carbon
nanotubes (CNTs) depend much on their geometrical structures, and
so can be easily changed by an applied pressure or strain, which
could be used to fabricate the nanoscale electromechanical
coupling devices and transducers. For example, a uniaxial strain
on the single-walled carbon nanotubes (SWNTs) can cause a
metal-semiconductor transition \cite{r1}.

Recently, the effect of hydrostatic pressure on the CNT bundle has
attracted much attention in experiments \cite{r2,r3, r4, r5, r6,
r7, r8, r9, r10}, including the Raman spectroscopy, and
theoretical calculations \cite{r11, r12, r13,r14}. The study of
the SWNT bundles indicates the Raman peaks shift to higher
frequencies with increasing hydrostatic pressure, and the radial
breathing mode (RBM) disappears from the spectrum above the
critical pressure, showing a SPT. On the other hand, there are
much fewer studies of the hydrostatic pressure's effect on the
DWNT bundle, which is the simplest multi-walled carbon nanotubes
(MWNTs). Very recently, the Raman measurements on the DWNT
materials under hydrostatic pressure \cite{r2,r3,r4} have found
the RBM intensity of the outer tubes decreases rapidly with
increasing pressure, exhibiting a similar behavior to that of SWNT
\cite{r5, r6, r7, r8, r9, r10, r11, r12, r13}, but the inner tubes
appear to be considerably less sensitive to the pressure, e.g.,
the pressure coefficient of the inner tube is 45\% smaller than
that of the outer tube.

Therefore, in this work, we intend to study the structural change
of DWNT bundles under hydrostatic pressure. Our numerical results
show that the symmetry and diameter of outer tube decide mainly
the change of cross section and response behavior of DWNT bundles
under the pressure. More importantly, two SPTs are found to exist
in some DWNT bundles if their outer tubes have the larger
diameters and no $C_{6}$ or $C_{3}$ symmetry.

The zero-temperature structural minimizations of the enthalpy
($H=U+PV$) were carried out on a supercell containing a
$2\times2\times2$ DWNT bundle using the universal force field
(UFF) method \cite{r15, r16}. In order to induce the SPT, a
step-wise increasing hydrostatic pressure was applied to the DWNT
bundle, minimizing the enthalpy of the DWNT bundle after each
pressure increment.

For (5,5)@(10,10), (7,7)@(12,12) and (9,0)@(18,0) DWNT bundles, we
also perform the first principles structure relaxations using the
total energy plane-wave potential method \cite{r17} in the
framework of local density approximation (LDA). The ion-electron
interaction was modeled by the projector augmented wave (PAW)
method \cite{r18} with a uniform energy cutoff of 500 eV. The
smearing width was taken to be 0.04 eV in the ground state. And a
very good agreement is obtained between the force field and the
first principles calculation results (see Table I), indicating
that the UFF method is suitable for the system researched in this
paper.

The transition pressures, $P_{c}$ and $P_{d}$, for different DWNT
bundles are listed in Table II. As we shall see, there exist three
kinds of the response behaviors to external pressure, and their
loading curves vs hydrostatic pressure are given in Fig.
1(a)-1(c), respectively.

Firstly, for small-diameter (5,5)@(10,10) DWNT bundle, a small
discontinuous volume change appears at $P=$ 18.01 GPa, accompanied
by a cross section's change between two deformed hexagons, as seen
from Fig. 1(a). At the same time, the inter tube (5,5) could not
collapse with a minimum distance of 4.35 {\AA} between its two
opposite walls, which is larger than the distance of 3.4 {\AA}
between nearby layers in the turbostratic graphite. If the
pressure is increased further, this distance will approach
continuously to 3.4 {\AA}. On the other hand, our simulation
indicates that the loading curve of (5,5) SWNT bundle [also shown
in Fig. 1(a)] varied with external pressure is continuous too and
no obvious SPT happens. But (10,10) SWNT bundle collapses at
$P_{d}=$ 3 GPa, forming a peanut-shaped cross section with a
separation of about 3.4 {\AA} between its two opposite parallel
walls. So, it is clear that existence of an inner (5,5) tube
increases the ability of the outer (10,10) tube to resist the
applied pressure, making so the (5,5)@(10,10) DWNT bundle do not
collapse at 18.01 GPa.

As for the intermediate-diameter DWNT bundles, e.g., the
(7,7)@(12,12), (9,9)@(14,14), (10,10)@(15,15), (11,11)@(16,16),
(12,12)@(17,17) and (13,13)@(18,18), as shown in Fig. 1(b), it is
found that all of them undergo one SPT and collapse completely.
Our simulations reveal similar collapses exist for the (7,7),
(9,9), (10,10), (11,11), (12,12) and (13,13) SWNT bundles. Taking
(10,10)@(15,15) DWNT bundle as an example, its collapse pressure
$P_{d}=$ 4.68 GPa that is higher than either that of (10,10) SWNT
bundle ($P_{d}=$ 3 GPa) or that of (15,15) SWNT bundle ($P_{d}=$
1.3 GPa). This means that the outer tube acts as a protection
shield, and the inner tube supports the outer one and increase its
structure stability. Our results are consistent with the
experimental measure results \cite{r2, r3, r4}.

However, the response behaviors of the larger-diameter DWNT
bundles become complex, as shown in Fig. 1(c). The (16,16)@(21,21)
and (19,19)@(24,24) DWNT bundles still collapse after one SPT.
Surprisingly, the (15,15)@(20,20), (17,17)@(22,22),
(18,18)@(23,23) and (21,21)@(26,26) DWNT bundles undergo two
different types of SPT with the transition pressure $P_{c}$ and
$P_{d}$ respectively, accompanied by their cross sections change
from deformed hexagon to racetrack first and then to peanut shape,
and the DWNT bundles collapse completely after the second SPT. It
should be noted that the DWNT bundles exhibit some small
discontinuous volume change between the different racetrack-shaped
cross sections. We think two SPTs and the racetrack-shaped cross
sections could be observed in future experiments if the pressure
increment is selected properly.

Furthermore, it is interesting to ask why the racetrack-shaped
cross sections do not appear for the (16,16)@(21,21) and
(19,19)@(24,24) DWNT bundles before their final collapse? Our
simulations reveal that not only their cross sections, but also
the ones of the intermediate-diameter DWNT bundles: (7,7)@(12,12),
(10,10)@(15,15), (13,13)@(18,18) remain relatively the better
hexagon with almost equal sides and corners just before the SPTs,
which are different from the other DWNT bundles with the deformed
hexagonal or racetrack cross sections before the SPT (in Fig.
1(b)-1(c) we only show the cross sections of some DWNTs, e.g., the
(9,9)@(14,14), (10,10)@(15,15), (13,13)@(18,18), (16,16)@(21,21)
and (18,18)@(23,23) as examples). It is known that the symmetry
group of the isolate (n,n) and (n,0) SWNT is $T_2^1 D_{nh}$. The
symmetries of $T_2^1$ and $\sigma _h $ are retained under
pressure, but in the bundle the cross section symmetry is reduced
from $D_{n}$ to $C_{6}$ for the specific outer tubes: (12,12),
(18,18) and (24,24), and to $C_{3}$ for the (15,15) and (21,21)
tubes. A necessary condition to form an ideal hexagonal lattice is
that the tube itself has a $C_{6}$ rotational axis, and $C_{3}$
symmetry can also be somewhat matched to the hexagonal lattice
symmetry. Thus, with increasing pressure, these outer tube's cross
section can keep better hexagon before the SPT. More importantly
these DWNT bundles only undergo one SPT to reach the stable
collapsed structure no matter how big is the diameter. On the
other hand, the DWNT bundles in which only their inner tubes have
the $C_{6}$ or $C_{3}$ symmetry, for example the (9,9)@(14,14),
(12,12)@(17,17), (15,15)@(20,20), (18,18)@(23,23) and
(21,21)@(26,26), have still the deformed hexagonal or
racetrack-shaped cross sections just before the SPT, and two SPTs
would happen with increased diameters. So, the symmetry of the
outer tube is a very important factor to decide the cross section
shape of the DWNT under the pressure, and comparatively the effect
of their inner tube is very small.

We have also studied the structure transformations of
zigzag@zigzag DWNT bundles under the hydrostatic pressure, which
are found to show a qualitatively same response mechanism to
external pressure with the armchair@armchair DWNT bundles, as
indicated by Table II.

Finally, in order to have an insight into the relationship of the
collapse pressure $P_{d}$ with the DWNT diameter and its tube
symmetry, we introduce the DWNT's average radius $R_{ave}$, which
is defined as an average value of the outer and inner tube's
radius of DWNT. Here we show the variation of $P_{d}$ with
$R_{ave}$ in Fig. 2, from which it can be seen that the collapse
pressure can be well fitted to $\sim 1/R_{ave}^{3}$ for the DWNT
bundles. It is known that the tube-tube coupling in a bundle is
described by the van der Waals force, which has a small effect on
the bundle's collapse pressure. And both the SWNT and DWNT can be
described by the continuum elasticity theory as the continuous
hollow cylinders. So their bundles would have similar response to
the hydrostatic pressure, and the only difference is that the
effective radius of DWNT is the average of inner and outer tubes'
radius. On the other hand, it has been proved by the continuum
elasticity theory [19] and the molecular dynamics simulations [20]
that the $P_{d}$ of an individual SWNT is inversely proportional
to its cubic radius. Thus, it is reasonable that the relation of
$P_{d}$ $\sim 1/R_{ave}^{3}$ appears in the DWNT bundle. In
addition, for the DWNT bundles in which the outer tube has $C_{6}$
or $C_{3}$ symmetry, the collapse pressure is larger than the
fitted result. This phenomenon is more obvious, especially for the
zigzag@zigzag DWNT bundles, e.g., the (9,0)@(18,0), (15,0)@(24,0),
(21,0)@(30,0), (27,0)@(36,0) and (33,0)@(42,0). The reason is that
both of their outer and inner tubes have $C_{6}$ or $C_{3}$
symmetry, further enhancing the matching to the hexagonal lattice
and increasing the structure stability of the system. This means
the matching between the DWNT symmetry and the lattice symmetry
can increase the ability of the DWNT bundle to resist the applied
pressure.

In addition, we have also made simulation on the (11,2)@(12,12)
DWNT bundle, which is composed of two coaxal SWNTs with different
chiral angles, and the result is shown in Table I, Fig. 1(b) and
2. The period of tube (11,2) is just seven times larger than that
of tube (7,7), and its radius is equal to that of (7,7). It can be
found from Fig. 1(b) that the (7,7)@(12,12) and (11,2)@(12,12)
DWNT bundles have the slightly different collapsed pressures of
$P_{d}=$ 10.6 GPa and 11.51 GPa, respectively, i.e. the chiral
symmetry of the inner tubes in the DWNT bundle has a smaller
effect on their collapse pressures.

In summary, our calculations show that the structural
transformations of DWNT bundles under hydrostatic pressure is
different from those of the SWNT bundles \cite{r5, r6, r7, r8, r9,
r10, r11, r12, r13} and isolate DWNT \cite{r21}. One or two SPTs
exist depending on the symmetry and diameter of DWNT bundles,
which manifest the complexity of nanotubes. It would be
interesting to experimentally determine mechanical (e.g.,
compressibility), optical (e.g., Raman spectrum) and thermal
(e.g., heat capacity) response functions of DWNT bundles to search
for signatures of these different types of structural
transformations.

\begin{acknowledgments}
This work was supported by the Natural Science Foundation of China
under Grant Nos. 10474035, 90503012, and 10304007, and also by the
State Key program of China through Grant No. 2004CB619004.
\end{acknowledgments}

\newpage


\begin{center}
\textbf{TABLE}
\end{center}

\begin{table}[htbp]
\caption{\label{tab1} Calculated lattice parameters of some DWNT
bundle without external pressure by UFF method and first
principles method. a, b and c are the lattice constants of DWNT
bundles, and $\alpha $, $\beta $ and $\gamma $ are the angles
between the two lattice vectors.}
\begin{ruledtabular}
\begin{tabular}{ccccccc}
DWNTs bundle& a& b& c& $\alpha $& $\beta $&
$\gamma $ \\
\hline (5,5)@(10,10)& 16.64& 16.64& 2.44& 85.79& 94.21&
120.16 \\
(5,5)@(10,10) LDA& 16.62& 16.62& 2.45& 85.78& 94.22&
120.09 \\
(7,7)@(12,12)& 19.26& 19.26& 2.44& 90.00& 90.00&
119.98 \\
(7,7)@(12,12) LDA& 19.26& 19.26& 2.45& 90.00& 90.00&
119.98 \\
(9,0)@(18,0)& 17.09& 17.09& 4.23& 89.19& 91.78&
120.01 \\
(9,0)@(18,0) LDA& 17.10& 17.09& 4.24& 89.18& 91.79&
120.01 \\
\end{tabular}
\end{ruledtabular}
\end{table}

\begin{table}[htbp]
\caption{\label{tab2} Calculated critical transition pressure
$P_{c}$ and $P_{d}$ of DWNT bundles. $R_{ave}$ is the average
value of the inner and outer tube's radius in the isolate DWNT
directly rolled up from the ideal graphene sheet. The suffix (NC)
means at this pressure, no collapse happens.}
\begin{ruledtabular}
\begin{tabular}{cccc}
DWNT bundle& $R_{ave}$& $P_{c}$& $P_{d}$\\
\hline (5,5)@(10,10)& 5.085& & 18.01 \par (NC) \\
(7,7)@(12,12)& 6.441& & 10.6 \\
(11,2)@(12,12)& 6.441& & 11.51 \\
 (9,9)@(14,14)& 7.797& & 5.1 \\
 (10,10)@(15,15)& 8.475& & 4.68\\
 (11,11)@(16,16)& 9.153& & 3.19 \\
 (12,12)@(17,17)& 9.831& & 2.39 \\
 (13,13)@(18,18)& 10.509& & 2.43 \\
 (15,15)@(20,20)& 11.865& 1.25& 1.44 \\
 (16,16)@(21,21)& 12.543& & 1.48 \\
 (17,17)@(22,22)& 13.221& 0.97& 1.1 \\
 (18,18)@(23,23)& 13.899& 0.7& 0.99 \\
 (19,19)@(24,24)& 14.577& & 1.02 \\
 (21,21)@(26,26)& 15.933& 0.32& 0.7 \\
\hline
(7,0)@(16,0)& 4.502& &30.5 \par (NC)\\
(9,0)@(18,0)& 5.284& & 20.24 \\
(11,0)@(20,0)& 6.067& & 12.12 \\
(13,0)@(22,0)& 6.85& & 7.69 \\
(15,0)@(24,0)& 7.633& & 6.61 \\
(17,0)@(26,0)& 8.416& & 4.42 \\
(19,0)@(28,0)& 9.199& & 2.99 \\
(21,0)@(30,0)& 9.982& & 3.61 \\
(23,0)@(32,0)& 10.765& & 1.82 \\
(25,0)@(34,0)& 11.548& 1.52& 1.55 \\
(27,0)@(36,0)& 12.331& & 2.23 \\
(29,0)@(38,0)& 13.113& 0.85& 1.13 \\
(31,0)@(40,0)& 13.896& 0.71& 0.99 \\
(33,0)@(42,0)& 14.679& & 1.01
\end{tabular}
\end{ruledtabular}
\end{table}

\newpage

\begin{figure}[htbp]
\includegraphics[width=0.8\columnwidth]{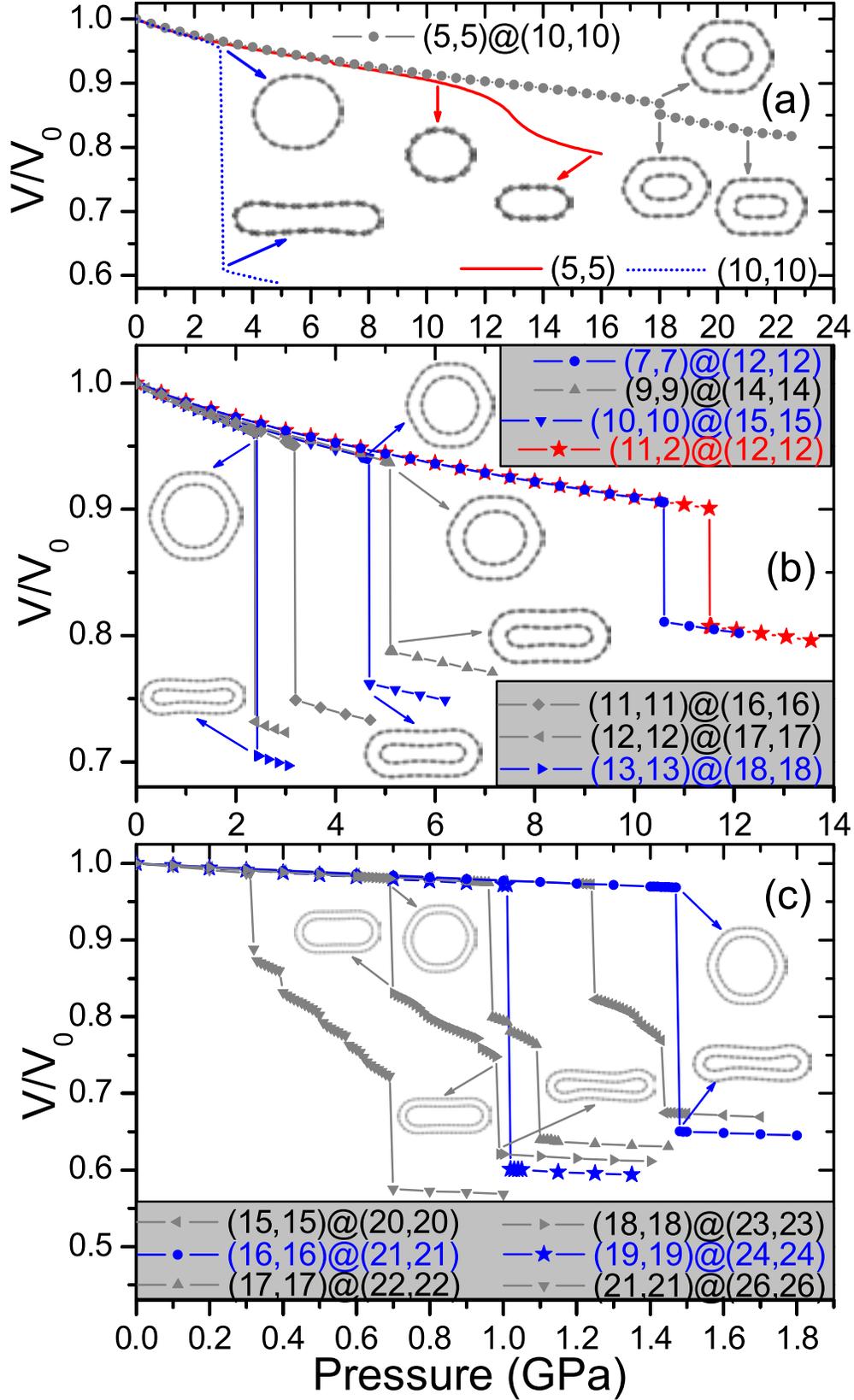}
\label{fig1} \caption{(Color online) Loading curves for different
armchair@armchair DWNT bundles as a function of hydrostatic
pressure. (a) small-diameter (5,5)@(10,10) DWNT bundle; (b) some
intermediate-diameter DWNT bundles; (c) the larger-diameter DWNT
bundles.}
\end{figure}

\begin{figure}[htbp]
\includegraphics[width=0.8\columnwidth]{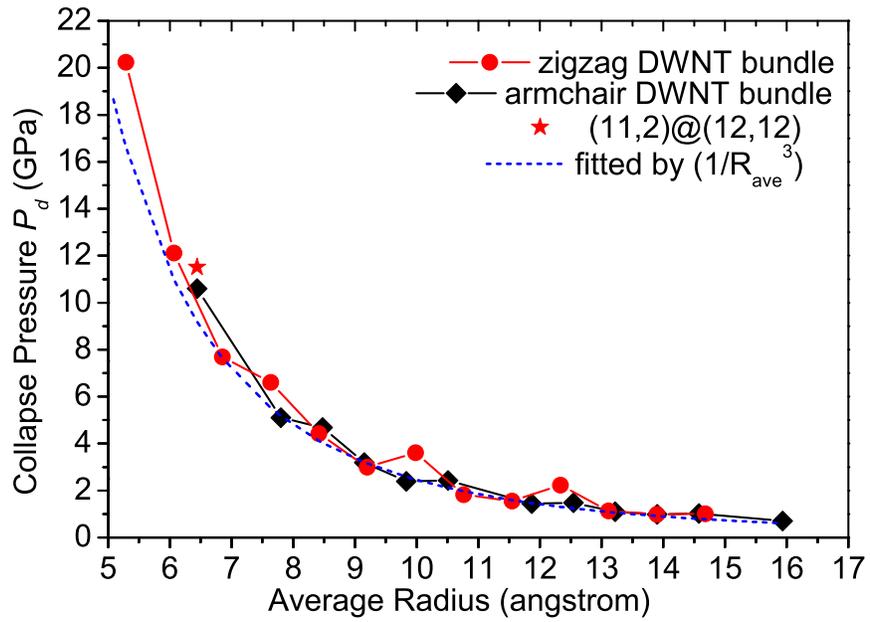}
\label{fig2} \caption{(Color online) Collapse pressure $P_{d}$ as a
function of the average radius of DWNT. $P_{d}$ can be well fitted
to $\sim 1/R_{ave}^{3}$.}
\end{figure}

\end{document}